\documentclass[prl,aps,showpacs,twocolumn,floatfix]{revtex4}
\usepackage{graphicx}

\begin{document}

\title{ Low momentum shell model effective interactions \\
with all-order core polarizations}

\author{Jason D.\ Holt$^{1}$, Jeremy W.\ Holt$^{1}$, T.\ T.\ S.\ Kuo$^{1}$, 
G.\ E.\ Brown$^{1}$, and S.\ K.\ Bogner$^{2}$}

\affiliation{$^1$Department of Physics, State Univeristy of New York at 
Stony Brook \\ Stony Brook, New York 11794, USA\\
$^2$Department of Physics, The Ohio State University, Columbus, Ohio 43210, 
USA}
\date{\today}

\begin{abstract}

An all-order summation of core polarization diagrams using the low-momentum 
nucleon-nucleon interaction $V_{\rm low-k}$ is presented. 
The summation is based on the Kirson-Babu-Brown (KBB) induced interaction 
approach in which the vertex functions are obtained self consistently by 
solving a set of non-linear coupled equations.
It is found that the solution of these equations is simplified by using 
$V_{\rm low-k}$, which is energy independent,
 and by employing Green functions in the 
particle-particle and particle-hole channels.
We have applied this approach to the $sd$-shell effective 
interactions and find that the results calculated to all orders using the 
KBB summation technique are remarkably similar to those of second-order
perturbation theory, average differences being less than 10$\%$.

\end{abstract}

\maketitle
\newpage

\noindent {\it Introduction.}--- 
Since the early works of Bertsch \cite{bertsch} and Kuo and Brown 
\cite{kuobrown}, the effect of core polarization in nuclear physics
has received much attention. Core polarization is particularly
important in the shell model effective interactions, 
where this process provides the long-range 
inter-nucleon interaction mediated by excitations of the core 
\cite{brownbook}. In microscopic calculations of effective
interactions, core polarization has played an essential role,
as illustrated by the familiar situation in $\rm ^{18} O$.
There the spectrum calculated with the bare $G$-matrix
was too compressed compared with experiment, while the 
inclusion of core polarization had the desirable effect 
of both lowering the $0^+$ ground state and raising the $4^+$ state, 
leading to a much improved agreement with experiment \cite{bertsch,kuobrown}.
As pointed out by Zuker \cite{zucker}, the Kuo-Brown matrix elements, 
although developed quite some time ago,
continue to be a highly useful shell model effective interaction. 
It should be noted that the core polarization 
(CP) diagrams associated with the above interactions were all 
calculated to second order (in the $G$-matrix) in perturbation theory. 
But what are the effects of core polarization beyond second order, and how 
can they be calculated? In this Letter we would like to address
these questions and present an all-order summation of CP diagrams for
the $sd$-shell interactions.

There have been a number of important CP studies beyond second order.
Third-order core polarization diagrams, including those with one fold,
were studied in detail by Barrett and Kirson \cite{barrettk} for the $sd$-shell
effective interactions.  Horth-Jensen {\it et al}.\ \cite{jensen95} have 
carried out extensive investigations of the third-order CP
diagrams for the tin region. A main result of these studies is that
the effect of the third-order diagrams is generally comparable to that
of the second order; the former cannot be ignored in comparison with
the latter. As is well known, high-order CP calculations are difficult to
perform, largely because the number of CP diagrams grows rapidly as one 
goes to higher orders in perturbation theory. The number of diagrams at third 
order is already quite large, though still manageable. Primarily because
of this difficulty, 
a complete fourth-order calculation has never been carried out. It was soon 
realized that an order-by-order calculation of CP diagrams beyond third order 
is not practicable. To fully assess the effects of core polarization to high
order, a non-perturbative method is called for.

In the present work, we shall use a non-perturbative method to carry out
an all-order summation of CP diagrams. Our method is based on 
the elegant and rigorous induced interaction approach of Kirson \cite{kirson}
and Babu and Brown \cite{babu}, hereafter referred to as the
KBB method. In this formalism the vertex functions are obtained
by solving a set of self-consistent equations, thereby generating
CP diagrams to all orders in a manner similar to the parquet summation 
\cite{jackson}. Using this approach, Kirson has studied 
$\rm ^{18}O$ and $\rm ^{18}F$ using a $G$-matrix interaction, and Jopko and 
Sprung \cite{jopko} have carried out a model study of this approach
using a separable interaction. A main conclusion of both studies is that
when CP diagrams are included to all orders the effective interaction
is very close to that given by the bare interaction alone. In contrast, 
Sj\"{o}berg \cite{sjoberg} applied the Babu-Brown formalism to nuclear matter 
and found that the inclusion of CP diagrams to all orders has 
a significant effect on the Fermi liquid parameters, in comparison with 
those given by the bare interaction. These conflicting results for CP studies
of finite nuclei and infinite nuclear matter have served as a primary
motivation for our present re-examination of the CP effect. 

Our application of the KBB formalism to shell-model effective interactions
is similar to that of Kirson, but our treatment is different in a number of 
important regards. As we will discuss, the particle-core and hole-core 
coupling
vertices used in the present work include a larger class of diagrams than have
been previously studied. We will show how the inclusion of these diagrams is
facilitated by using the recently 
developed low-momentum nucleon-nucleon interaction $V_{\rm low-k}$
\cite{bogner02,kuorg02,cora02,cora02b,bogner03,cholt04,fholt04} instead of
the previously used $G$-matrix. This is primarily because the
$G$-matrix \cite{krenci,jensen95} depends on both starting energy 
and Pauli exclusion operator, while $V_{\rm low-k}$ depends on neither.
It is noted that the S-wave interactions calculated from the 
Moszkowski-Scott separation method gave essentially the same matrix elements
as $V_{\rm low-k}$.\cite{jeremy}
In the subsequent Formalism section we will discuss these topics in greater
detail, and in the final section we present our results together with a 
summary.
Although the present calculation is restricted to the $sd$-shell
effective interactions, the methods we develop can be readily expanded to
other nuclear regions and with appropriate generalization applied to
effective operators such as magnetic moments.
\vskip.1in

\begin{figure}
\includegraphics[height=2.2in]{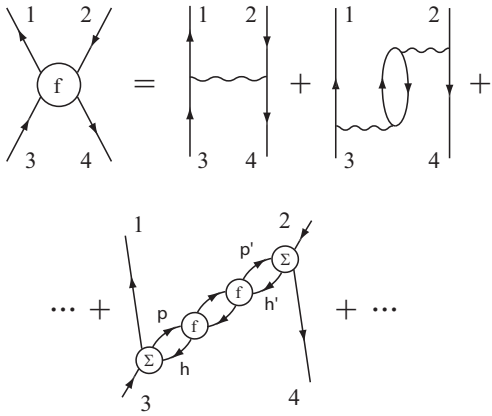}
\caption{Self-consistent diagramatic expansion of the $ph$ vertex 
function, $f$, 
where $\Sigma$ is defined in the text.} 
\label{fvert}
\end{figure}

\begin{figure}
\includegraphics[height=1.5in]{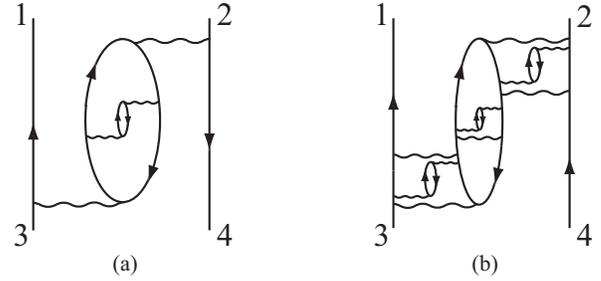}
\caption{Higher-order terms contributing to the vertex functions $f$ and
$\Gamma$, including (a) nested bubbles in bubbles and (b) particle-core and
hole-core couplings.} 
\label{gkb}
\end{figure}

\noindent {\it Formalism.}--- 
The KBB induced interaction approach provides a very appealing way
for summing up planar diagrams to all orders.
Its  fundamental requirement is that the irreducible vertex functions be 
calculated self-consistently. This means that any core polarization term 
contained in a vertex function must be generated self-consistently  
from the same vertex function. To illustrate this point, we consider the 
particle-hole ($ph$) vertex function $f$. As shown in 
\mbox{Fig.\ \ref{fvert}},
$f$ is generated by summation of the driving term $V$ 
and core polarization terms, the latter being dependent on $f$.  
This then gives the self-consistent equation for $f$
\begin{equation}
f=V+\Sigma g_{ph}\Sigma+\Sigma g_{ph}fg_{ph}\Sigma
+\Sigma g_{ph}fg_{ph}fg_{ph}\Sigma+\cdots,
\label{f1}
\end{equation}
where $g_{ph}$ is the free $ph$ Green function, and $\Sigma$ denotes the vertex
for particle-core and hole-core coupling. 
The second-order CP diagram of Fig.\ 1 is the lowest-order term
contained in $\Sigma g_{ph} \Sigma$.  We note that for simplicity 
the bra and ket indices have been suppressed in the above 
equation as well as the equations that follow. 
For example, in \mbox{Eq.\ (\ref{f1})} the $f$ on the LHS represents 
$\langle 12^{-1}|f|34^{-1} \rangle$, while the fifth and sixth $\Sigma$'s
on the RHS represent $\langle 1ph^{-1}|\Sigma|3 \rangle$ and 
$\langle 2^{-1}|\Sigma|p'h'^{-1}4^{-1} \rangle$, respectively.
To further visualize the structure of $f$, let us consider 
$\Sigma = V$ in Eq.\ (\ref{f1}):
\begin{equation}
f=V+Vg_{ph}V+Vg_{ph}fg_{ph}V+Vg_{ph}fg_{ph}fg_{ph}V+\cdots.
\end{equation}
Since $f$ appears on both sides of this equation, it is clear that an iterative
solution for $f$ will yield core polarization diagrams to all orders, 
including those with ``bubbles inside bubbles'', like those shown in diagram 
(a) of \mbox{Fig.\ \ref{gkb}}.

For nuclear many-body calculations in general, we also need the 
particle-particle ($pp$) vertex function $\Gamma$. Like $f$, $\Gamma$ is given 
by a driving term plus core polarization terms. Furthermore, the diagramatic 
representation of $\Gamma$ is identical to \mbox{Fig.\ \ref{fvert}} except that
the hole lines 2 and 4 are replaced by corresponding particle lines.
This gives the self-consistent equation for $\Gamma$
\begin{equation}
\Gamma=V+\Sigma g_{ph}\Sigma+\Sigma g_{ph}fg_{ph}\Sigma
+\Sigma g_{ph}fg_{ph}fg_{ph}\Sigma+\cdots.
\label{gam1}
\end{equation}
To clarify our compact notation, we note that the external lines of the 
$\Sigma$
vertices in $\Gamma$ are different than those shown in Fig.\ \ref{fvert}. The
upper $\Sigma$ vertex, for example, now represents 
$\langle 2 | \Sigma | ph^{-1}4
\rangle$. These different $\Sigma$ vertices can be related to each other,
however, via appropriate particle-hole transformations.

Finally, the vertex functions $f$ and $\Gamma$ are coupled together via the 
coupling vertex $\Sigma$. In the present work we choose
\begin{equation}
\Sigma = V+\Sigma_{ph}+\Sigma_{pp}, \label{sig1}
\end{equation}
\begin{equation}
\Sigma_{ph} = Vg_{ph}V+Vg_{ph}fg_{ph}V
            + Vg_{ph}fg_{ph}fg_{ph}V + \cdots, \label{sigph}
\end{equation}
\begin{equation}
\Sigma_{pp} = V g_{pp}V +V g_{pp} \Gamma g_{pp} V
              +V g_{pp} \Gamma g_{pp}\Gamma g_{pp} V+\cdots, \label{sigpp}
\end{equation}
where $g_{pp}$ is the free $pp$ Green function.

The self-consistent vertex functions $f$ 
and $\Gamma$ are determined from \mbox{Eqs.\ (\ref{f1}) and
(\ref{gam1}--\ref{sigpp})}.  These are similar to the 
equations used by Kirson \cite{kirson}, except that our $\Sigma$ 
includes both $\Sigma_{ph}$ and $\Sigma_{pp}$, while the equivalent term in 
Kirson's calculations only includes $\Sigma _{ph}$ \cite{kirson,ellisrmp}. 
To see the role of the $\Sigma$ vertices,
let us consider diagram (b) of \mbox{Fig.\ \ref{gkb}}. 
Here the lower particle-core
vertex, which contains repeated particle-particle interactions, belongs to 
$\Sigma _{pp}$, while the upper one, which contains repeated particle-hole 
interactions, belongs to $\Sigma_{ph}$. 
It is, of course, necessary to include $\Sigma _{pp}$ in order to have such 
CP diagrams in the all-order sum. Our equations are equivalent to those
of Kirson when $\Sigma _{pp}$ is set to zero, and in this case 
$\Gamma$ does not enter the calculation of $f$.

Solving the above equations for $f$ and $\Gamma$ may seem complicated, but
we have found their solution can be simplified significantly 
through use of the true $ph$ and $pp$ Green functions
\begin{equation}
G_{ph} = g_{ph}+g_{ph}fG_{ph},\label{fullgfph}
\end{equation}
\begin{equation}
G_{pp} = g_{pp}+g_{pp}\Gamma G_{pp}. \label{fullgfpp}
\end{equation}
Using these Green functions to partially sum and regroup our series,
the self-consistent \mbox{Eqs.\ (\ref{f1}) and (\ref{gam1}--\ref{sigpp})} 
assume a much simpler form 
\begin{equation}
f = V+\Sigma G_{ph}\Sigma,\label{fgf}
\end{equation}
\begin{equation}
\Gamma = V+\Sigma G_{ph}\Sigma, \label{gammagf}
\end{equation}
\begin{equation}
\Sigma = V+VG_{ph}V +VG_{pp}V. \label{vert2}
\end{equation}
The above simplifications also aid our numerical efforts, and using the 
following iterative method we find our coupled equations can be solved rather 
efficiently.
For the $n^{th}$ iteration, we start from $f^{(n)}$ and $\Gamma ^{(n)}$,
to first calculate $G_{ph}^{(n)}$ and $G_{pp}^{(n)}$ followed by 
$\Sigma ^{(n)}$, as seen from \mbox{Eqs.\ (\ref{fullgfph}--\ref{vert2})}.
The vertex functions for the subsequent iteration are then obtained 
by taking $f^{(n+1)}=V+\Sigma^{(n)} G_{ph}^{(n)}\Sigma^{(n)}$ and 
$\Gamma^{(n+1)}=V+\Sigma^{(n)} G_{ph}^{(n)}\Sigma^{(n)}$.
The entire iterative process begins from the initial 
$f^{(0)}=V+Vg_{ph}V$ and $\Gamma^{(0)}=V+Vg_{ph}V$, and typically converges 
after just a few iterations.

In the present work, we have included folded diagrams to all orders.
As detailed in \cite{ko90}, we use this method to reduce the full-space 
nuclear many-body problem $H\Psi _n=E_n \Psi _n$ to a model space
problem $H_{\rm eff}\chi _m=E_m \chi _m$, where $H=H_0+V$,
$H_{\rm eff}=H_0+V_{\rm eff}$ and $V$ denotes the bare NN interaction.
The effective interaction $V_{\rm eff}$ is given by the folded-diagram
expansion
\begin{equation}
V_{\rm eff} = \hat{Q} - \hat{Q'} \int \hat{Q} + \hat{Q'} \int \hat{Q} \int 
\hat{Q} -  \cdots.
\label{veff}
\end{equation}
We consider the effective interactions for valence nucleons in the $sd$-shell,
and in this case $\hat Q$ is given by the 
vertex function $\Gamma$ obtained earlier from the KBB equations. 
In \cite{kirson}, the effect of higher order CP diagrams to
the non-folded $\hat Q$ term was extensively studied. In the present work,
we first calculate $\hat Q$ including CP diagrams to all orders. Then the
above folded diagram series for $V_{\rm eff}$ is summed to all orders 
using the Lee-Suzuki iteration method \cite{jensen95}, explicitly including 
folded CP diagrams to all orders. 

\begin{figure}
\includegraphics[height=2.5in, angle=270]{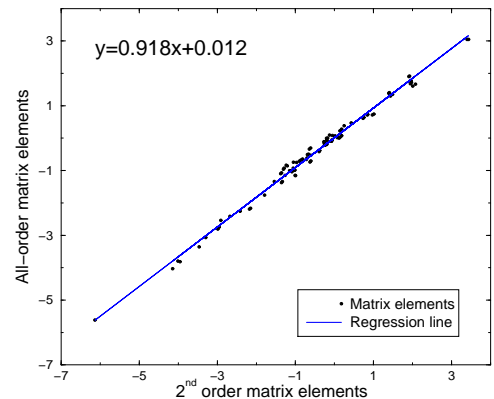}
\caption{A comparison of the second-order core 
polarization matrix elements with those of the all-order KBB calculation.} 
\label{fig.3}
\end{figure}

For the present calculation we have chosen to use the low-momentum 
nucleon-nucleon interaction, $V_{\rm low-k}$. Since the vertex functions
$f$ and $\Gamma$ both depend on the starting energy, there would be 
off-energy-shell effects present in many CP diagrams if the
$G$-matrix interaction were chosen. This would make the calculation very
complicated. $V_{\rm low-k}$, on the other hand, is energy independent so no 
such difficulties are encountered.
Since detailed treatments of $V_{\rm low-k}$ have been given elsewhere
\cite{bogner02,kuorg02,cora02,cora02b,bogner03,cholt04,fholt04}, 
here we provide only a brief description.
We define $V_{\rm low-k}$ through the $T$-matrix equivalence equations

\begin{eqnarray}
T(k',k,k^2) = V_{NN}(k',k) + P\int _0 ^{\infty} q^2 dq V_{NN}(k',q) 
\nonumber \\
\times \frac{1}{k^2-q^2} T(q,k,k^2 )
\end{eqnarray}
\begin{eqnarray}
\lefteqn{T_{low-k }(p',p,p^2)  = V_{low-k }(p',p)}  \\
&&  + P\int _0 ^{\Lambda} q^2 dq  V_{low-k }(p',q) \nonumber
 \frac{1}{p^2-q^2} T_{low-k} (q,p,p^2)
\end{eqnarray}
\begin{eqnarray}
T(p',p,p^2) &=& T_{low-k}(p',p,p^2); ~( p',p) \leq \Lambda,
\end{eqnarray}

where $V_{NN}$ represents some realistic NN potential and $\Lambda$ is the 
decimation momentum beyond which the high-momentum components of $V_{NN}$ 
are integrated out. $V_{\rm low-k}$ preserves both the deuteron binding energy 
and the low-energy scattering phase shifts of $V_{NN}$.
 Since empirical nucleon scattering
phase shifts are available only up to the pion production threshold 
($E_{lab}\sim \rm350 \ MeV$), beyond this momentum the realistic NN potentials
cannot be uniquely determined. Accordingly, we choose 
$\Lambda \approx 2.0$ fm$^{-1}$ thereby retaining only the information from a 
given potential that is constrained by experiment.
In fact for this $\Lambda$, the $V_{\rm low-k}$ derived from various
NN potentials
\cite{cdbonn,argonne,nijmegen,idaho} are all nearly identical \cite{bogner03}.
Except where noted otherwise, in our calculations we employ 
the $V_{\rm low-k}$ derived from the CD-Bonn potential \cite{cdbonn}.

\begin{figure}
\includegraphics[height=3.2in,angle=270]{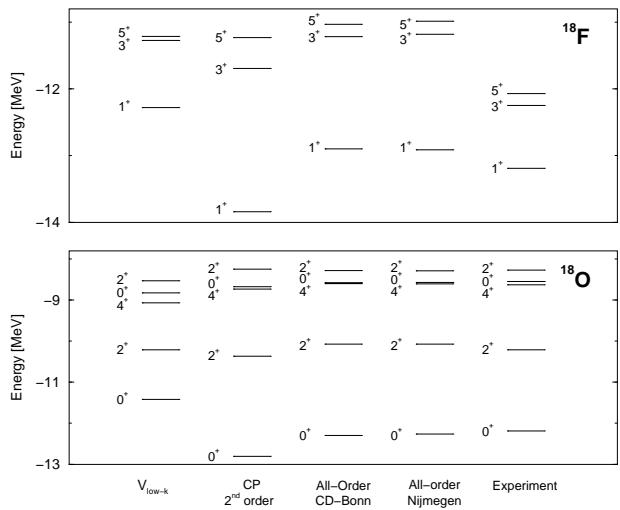}
\caption{Spectra for the $^{18} \rm F$ (top) and $^{18} \rm O$ 
(bottom) systems calculated to different orders in perturbation theory.} 
\label{fig.4}
\end{figure}

\noindent {\it Results and Discussion.}--- 
As an initial study, we have carried out a restricted all-order
CP calculation for the $sd$-shell effective interactions. In particular, 
we sum only the TDA diagrams for the Green functions $G_{pp}$ 
and $G_{ph}$, leaving a study of RPA diagrams to a future
publication. 
Also, we use a limited shell model space consisting of 10 
orbits from $0s_{1/2}$ to $1p_{1/2}$ and an oscillator constant of 
$\hbar \omega$=14 MeV. Only core excitations within this space are 
included. Vary, Sauer and Wong \cite{vary} have pointed out that
for CP diagrams one needs to include intermediate states of high excitation 
energies (up to $\sim 10 \hbar \omega$) in order for the second-order
core polarization term to converge. In their work a $G$-matrix derived
from the Reid soft-core potential was used, but our use of $V_{\rm low-k}$ 
may yield different results as it has greatly reduced 
high-momentum components.  We plan to 
study this convergence problem for $V_{\rm low-k}$ in the near future.

With these restrictions, we have calculated $V_{\rm eff}$ from Eqs.\ 
(\ref{fullgfph}--\ref{veff}).
A large number of angular momentum recouplings are involved
in calculating the CP diagrams. In this regard, we have followed
closely the diagram rules in \cite{kuoshur}. In Fig.\ \ref{fig.3}
we compare the $sd$-shell $V_{\rm eff}$ matrix elements calculated
with all-order CP diagrams with those from second-order \cite{bogner02}
CP diagrams. A least-squares fit was applied to the data, and 
it is apparent that the effect of including CP to all orders in our 
calculation is a mild suppression of the second-order contributions.
This conclusion is further born out in the calculation of the 
$\rm ^{18}O$ and $\rm ^{18}F$ spectra, the results of which are shown 
in Fig.\ \ref{fig.4}. Here we observe a weak suppression of the
second-order effects in $\rm ^{18}O$ but a moderate suppression in 
$\rm ^{18}F$. In the same figure we also observe that the spectra for 
different $V_{\rm low-k}$ derived from the CD-Bonn and Nijmegen bare 
potentials are nearly identical.

In summary, we have presented a method based on the KBB induced interaction
formalism for efficiently summing core polarization diagrams to all orders 
in perturbation theory. This summation is carried out by way of
the KBB self-consistent equations whose solution is significantly
simplified by the use of the true $pp$ and $ph$ Green functions,
and by the use of the energy-independent $V_{\rm low-k}$.
Although our calculation was restricted in several
important aspects, we find that our final renormalized interaction 
is remarkably close to that of second-order perturbation theory. This is of 
practical importance and a welcoming result, for it allows one to use 
the results from a second-order
calculation to approximate the contributions resulting from a large class of 
higher-order diagrams. In the future we intend both to expand our treatment by 
including additional diagramatic contributions (RPA) and to use the results in 
calculations of other nuclear observables.

\begin{acknowledgments}
We thank M.\ Kirson for helpful discussions. 
Partial support from the US Department of Energy under contracts
DE-FG02-88ER40388 is gratefully acknowledged.
\end{acknowledgments}

\end{document}